\documentclass{article}

\usepackage{amsmath}  
\usepackage[english]{babel}
\usepackage[utf8]{inputenc}
\usepackage{amssymb}
\usepackage{graphicx}
\usepackage{authblk}
\usepackage{caption}
\usepackage{multirow}
\usepackage{amscd}
\usepackage{geometry}
\usepackage{a4wide}
\usepackage[usenames]{color}
\usepackage{etaremune} 
\usepackage[hidelinks]{hyperref}
\usepackage{amsfonts}
\usepackage{ifthen}
\usepackage{color}
\usepackage{amscd}
\usepackage{latexsym}
\usepackage{mathtools}
\usepackage{subfigure}
\usepackage{amsmath}
\usepackage{amsbsy}

\newcommand{\bd}{\begin{definition}}
	\newcommand{\ed}{\end{definition}}
\newcommand{\bt}{\begin{theorem}}
	\newcommand{\et}{\end{theorem}}
\newcommand{\bi}{\begin{itemize}}
	\newcommand{\ei}{\end{itemize}}
\newcommand{\ben}{\begin{enumerate}}
	\newcommand{\een}{\end{enumerate}}
\newcommand{\beq}{\begin{equation}}
\newcommand{\eeq}{\end{equation}}

\newcommand{\R}{\mbox{$ \mathbb{R}  $}}
\newcommand{\C}{\mbox{$ \mathbb{C} $}}

\newcommand{\h}{\mathcal{H}}

\newcommand{\half}{\frac{1}{2}} 
\newcommand{\pmat}[1]{\begin{pmatrix} #1 \end{pmatrix}}

\newtheorem{definition}{Def.}[section]
\newtheorem{theorem}{Theorem}[section]

\def\keywords{\vspace{.5em}
	{\noindent \textbf{Keywords}:\,\relax%
}}

\begin{document}

\title{The classification of rebit quantum channels}

\author[1]{Michele Aldé\thanks{michele.alde@studenti.unimi.it}}
\author[2]{Michel Berthier\thanks{michel.berthier@univ-lr.fr}}
\author[3]{Edoardo Provenzi\thanks{edoardo.provenzi@math.u-bordeaux.fr}}
\affil[1]{Università degli studi di Milano, Dipartimento di matematica, via Saldini 50, 20133 Milano, Italy}
\affil[2]{Laboratoire MIA, Batiment Pascal, Pôle Sciences et Technologie, Université de La Rochelle, 23, Avenue A. Einstein, BP 33060, 17031 La Rochelle cedex, France}
\affil[3]{Université de Bordeaux, CNRS, Bordeaux INP, IMB, UMR 5251\\ F-33400, 351 Cours de la Libération, Talence, France}

\renewcommand\Authands{ and }
\date{}

\maketitle


\begin{abstract}
The classification of qubit channels is known since 2002. However, that of rebit channels has never been studied so far, maybe because of the scarcity of concrete rebit examples. In this paper we point out that the strategy used to classify qubit channels cannot be pursued in the rebit case and we propose an alternative which allows us to complete the rebit channel classification. This mathematical result has not only a purely abstract interest: as we shall briefly mention, it may have applications in the analysis of local properties and temporal evolution of real quantum systems and also in a recent color vision model based on quantum information. 
\end{abstract}

\keywords{Rebit channels, quantum information, effects.}

Quantum channels represent the most general transformations between states of interacting, or open, quantum systems. Their classification is a complicated research topic and can only be achieved in very special cases. An example of paramount importance is represented by qubit channels, whose classification has been exhibited in \cite{Ruskai:2002}. 

Up to the authors' knowledge, the analogous result for rebit systems has never been obtained. In this paper we start by showing that the technique developed in \cite{Ruskai:2002}, based on Choi's theorem, cannot be applied in the rebit case. We then propose an alternative strategy, based on the $\chi$-map representation, which permits to achieve the complete rebit channel classification. 

While qubits are ubiquitous in quantum information, rebits remained for many decades a very interesting yet completely abstract research topic, see for instance \cite{Wootters:2014}. This is likely to be the cause underlying the lack of interest in the rebit channel classification. For this reason, we deem important to underline the practical relevance of the classification theorem by briefly mentioning possible applications in the analysis of local and dynamical features of rebit systems and also in framework offered by a recent quantum-like color perception theory. These applications will be developed in separate future works. 



The plan of the paper is the following: we start in section \ref{sec:mathpreliminars} by briefly recalling the basic terminology and mathematical results about open quantum systems, then we pass to the rebit channel classification in section   \ref{sec:classification}, which will be subdivided in several subsection for the sake of a smoother reading. In section \ref{sec:noticeablechannels} we will show how to single out some noticeable rebit channels using the classification theorem. Finally, possible concrete applications of our mathematical results will be discussed in section \ref{sec:applications}. 

\section{Terminology and preliminary mathematical results about open quantum systems}\label{sec:mathpreliminars}

Here we introduce the terminology and the known mathematical results of the theory of composite quantum systems that will be used in the rest of the paper. Unless otherwise stated, the main reference for all the results that will be quoted is \cite{Heinosaari:2011}. Since our aim is to study rebit quantum channels, all the Hilbert spaces that we will consider are implicitly assumed to be \textit{finite-dimensional} and \textit{real}.

Given a quantum system $S$, if $A$ and $B$ are two identifiable parts of it, i.e. two subsystems on which we are able to perform experiments to address the properties of $A$ and $B$ individually, then we say that $S$ is the composite system of $A$ and $B$ and we write $S=A+B$. If $\h_A$ and $\h_B$ are the Hilbert state spaces underlying the systems $A$ and $B$, respectively, then the quantum state space of $A+B$ is $\h_A \otimes \h_B$. 

If $A$ is not isolated, i.e. it can interact with $B$, then we say that $A$ is an open quantum system and $B$ is its environment. If $L(\h_A)$ denotes the vector space of linear operators on $\h_A$, then a linear map $C:L(\h_A) \to L(\h_A)$ is said to be a \textit{channel} if it is trace preserving and completely positive. This last request corresponds to demand not only that $C$ is positive, i.e. it preserves the positivity\footnote{$T\in L(\h_A)$ is positive if, for all $x\in \h_A$, $\langle x,T x\rangle \ge 0$, where $\langle \; , \; \rangle$ is the inner product of $\h_A$.} of operators on $\h_A$, but also that its extension, i.e. the map  $C \otimes id_B:L(\h_A\otimes \h_B)\to L(\h_A\otimes \h_B)$, is positive for all $B$. The requests defining a channel $C$ are the minimal ones to guarantee that $C$ maps states in states and that $C$ does not introduce non-meaningful negative probabilities when experiments on the composite quantum system $A+B$ are performed. To simplify the notation, let us denote from now on $\h_A$ as $\h$. 

An important class of channels is represented by the orthogonal ones: given an orthogonal operator $O\in L(\h)$, the \textit{orthogonal channel} associated to $O$ is the map $\sigma_O:L(\h) \to L(\h)$ such that, for all $T\in L(\h)$, $\sigma_O(T):= O\, T\, O^t$. Orthogonal channels are \textit{unital}, i.e. they map the identity into itself, moreover, they form a group w.r.t. functional composition. 

For the classification of channels their matrix representations turn out to be extremely useful. The most basic of these representations can be obtained by 
equipping $L(\h)$ with the Hilbert-Schmidt (HS) product, i.e. $\langle T,S\rangle_{\rm HS}={\rm Tr}(S^tT)$ for all $T,S\in L(\h)$, and fixing an orthonormal basis $(E_j)_{j=0}^{d^2-1}$ of operators of $L(\h)$. This permits to decompose $T\in L(\h)$ as follows
\beq 
T=\sum_{j=0}^{d^2-1} t_j^T E_j = \sum_{j=0}^{d^2-1} {\rm Tr}(E_j^t T) E_j
\eeq 
and to uniquely associate it to the vector ${\bf t}^T=(t^T_j)_{j=0}^{d^2-1}\in \R^{d^2}$. If $C:L(\h)\to L(\h)$ is a channel, then, by linearity, it follows that 
\beq 
C(T)=\sum_{j=0}^{d^2-1} t^{C(T)}_j E_j,
\eeq 
where
\beq\label{eq:tildet}
t^{C(T)}_j = \sum_{k=0}^{d^2-1} C_{jk} t^T_k \quad {\rm and } \quad C_{jk}={\rm Tr}(E_j^t \, C(E_k)).
\eeq
${\bf t}^{C(T)}=(t^{C(T)}_j)_{j=0}^{d^2-1}$ is the vector in $\R^{d^2}$ uniquely associated to $C(T)$ w.r.t. the basis $(E_j)_{j=0}^{d^2-1}$. 

A more sophisticated representation than the one just introduced is the so-called  \textit{$\chi$-matrix representation}, which will turn out to be a key ingredient for the rebit channel classification. This representation involves the so-called space of `super-operators', i.e. the $(d^2\times d^2)$-dimensional vector space $L((L(\h))$ of linear maps from $L(\h)$ to itself endowed with the following inner product inherited by the Hilbert-Schmidt product on $L(\h)$:
\beq
\langle \Psi, \Phi\rangle:=\sum_{j=0}^{d^2-1}{\langle\Psi(E_j),\Phi(E_j)\rangle_{\rm HS}},\qquad \forall \Psi,\Phi\in L(L(\h)).
\eeq
In the rebit channel classification, we will be interested in the action of channels on density matrices, which are symmetric, hence it is more useful to replace $L(L(\h))$ with $L({\rm Sym}(\h))$, where ${\rm Sym}(\h)\subset L(\h)$ is the vector subspace of dimension $N=d(d+1)/2$ of symmetric linear operators on $\h$. 

Given an orthonormal basis $(E_j)_{j=0}^{N-1}$ of operators of ${\rm Sym}(\h)$, an orthonormal basis $(F_{rs})_{r,s=0}^{N-1}$ for $L({\rm Sym}(\h))$ is defined by  
\beq
F_{rs}(T):=E_r T E_s^t, \qquad \forall T\in {\rm Sym}(\h),
\eeq 
so we can decompose a channel $C$ on $(F_{rs})_{r,s=0}^{N-1}$ as
\begin{equation}\label{eq: decomposition C}
	C=\sum_{r,s=0}^{N-1}\chi_{rs}F_{rs},
\end{equation}
where the coefficients $\chi_{rs}$ are obtained via inner product of $C$ with the basis elements $F_{rs}$ and can be written explicitly as follows
\beq\label{eq: coefficients chi}
\chi_{rs}=\sum_{j,k=0}^{N-1}C_{jk}{\rm Tr}(E_j E_sE_k^t E_r^t),
\eeq
with $C_{jk}$ as in eq. (\ref{eq:tildet}). The $\chi$-matrix uniquely associated to $C$ is  $\chi_C:=(\chi_{rs})_{r,s=0}^{N-1}$ and it has the property that \textit{$C$ is completely positive if and only if $\chi_C$ is positive semi-definite}, see \cite{Heinosaari:2011} page 206.

The last matrix representation that we need to recall is the \textit{Bloch representation}, which relies on the choice of a \textit{Bloch basis} $E_0:=I_d\cup \vec E$, $\vec E:=(E_1,\dots,E_{N-1})$, i.e. an orthogonal basis (w.r.t. the Hilbert-Schmidt product) of symmetric traceless operators on $\h$ such that $\|E_k\|^2_{\rm HS}=\|E_0\|^2_{\rm HS}=d$ for all $k=1,\dots,N-1$. The corresponding \textit{Bloch decomposition} of $T$ is written as follows:
\beq\label{eq:Blochdecomp} 
T = \frac{1}{d} ({\rm Tr}(T)I_d+{\bf v}^T \cdot \vec E) = \frac{1}{d}({\rm Tr}(T)I_d + \sum_{k=1}^{N-1}  v_k^T E_k), 
\eeq 
where ${\bf v}^T\in \R^{N-1}$ is called \textit{Bloch vector} and has components $v_k^T={\rm Tr}( E_kT)$, $k=1,\dots,N-1$. 

The relationship between the vector ${\bf t}^T$ and the Bloch vector ${\bf v}^T$ is the following
\beq\label{eq:trhov}
{\bf t}^T=\frac{1}{d}({\rm Tr}(T),{\bf v}^T).
\eeq 
Given a channel $C$, the relationship between the vector $\textbf{t}^{C(T)}$ and the Bloch vector ${\bf v}^{C(T)}$ becomes (see \cite{Heinosaari:2011} page 205)
\beq\label{eq:vblochannel}
{\bf t}^T=\frac{1}{d}({\rm Tr}(T),{\bf v}^{C(T)}), \quad {\rm with} \quad  {\bf v}^{C(T)} = {\bf w} + A{\bf v}^T,
\eeq 
where  the components of the vector ${\bf w}\in \R^{N-1}$ and the entries of the matrix $A\in M(N-1,\R)$ are
\beq
w_j={\rm Tr}(T)C_{j0}, \quad A_{jk} = C_{jk}, \quad j,k=1,\dots,N-1.
\eeq 
Two features of the Bloch representation can be singled out: after the action of a channel, \textit{the first component of} ${\bf t}^T$ \textit{remains invariant}, while the \textit{the Bloch vector undergoes an affine transformation}. 

For later purposes, we remark a third property:  \textit{whenever a linear map leaves the first component of} ${\bf t}^T$ \textit{invariant}, for all $T$, \textit{that map is trace-preserving}.

For orthogonal channel we can say more, as stated in the following theorem, see \cite{Heinosaari:2011} page 205.

\bt\label{th:roto}
If $C$ is a orthogonal channel $\sigma_O$, then 
\beq 
{\bf v}^{\sigma_O(T)}=R_O{\bf v}^T \, , \qquad R_O\in \emph{SO}(N-1), 
\eeq 
with:
\beq\label{eq:oro}
(R_O)_{jk} = \frac{1}{d}{\rm Tr}(E_jOE_k O^t), \quad j,k=1,\dots,N-1, 
\eeq 
i.e. the Bloch vector of $T\in L(\h)$ undergoes a rotation after the application of an orthogonal channel:
\beq\label{eq:unitary channel}
\sigma_O(T)=\frac 1 d(I_d+R_O{\bf v}^T\cdot\vec{E}).
\eeq 
\et


\section{Classification of rebit channels}\label{sec:classification}

In this section we will carry out the complete classification of rebit channels. Due to the length of this procedure, we have subdivided  the section in several subsections to facilitate the reading. 

\subsection{The first step: decomposition of a rebit channel in the orthogonal and diagonal parts}
To study the classification of rebit channels we fix $\h=\R^2$ and the Bloch basis $(\sigma_j)_{j=0}^2$, where $\sigma_0$ is $I_2$ and $\sigma_1,\sigma_2$ are the (symmetric and traceless) real Pauli matrices:
\begin{equation}
	\sigma_1=\begin{pmatrix}
		1 & 0 \\ 0 & -1
	\end{pmatrix}, \; \sigma_2=\begin{pmatrix}
		0 & 1 \\ 1 & 0
	\end{pmatrix}.
\end{equation} 
Moreover, instead of considering a general operator $T$, we will deal directly with density matrices, i.e. positive unit-trace operators $\rho\in L(\R^2)$ representing rebit states. From eq. \eqref{eq:Blochdecomp}, we obtain immediately that the Bloch decomposition of $\rho$ is
\beq
\rho=\frac{1}{2}(\sigma_0+{\bf v}^\rho \cdot \vec \sigma) \equiv \rho_0 + s_1 \sigma_1 + s_2 \sigma_2,
\eeq 
where $\rho_0=I_2/2$ represents the maximally mixed state and ${\bf v}^\rho=(s_1,s_2)^t\in \cal D$ is the Bloch vector associated to $\rho$, $\mathcal D$ being the unit disk in $\R^2$.

\begin{figure}[ht!]
	\centering
	\includegraphics[scale=0.9]{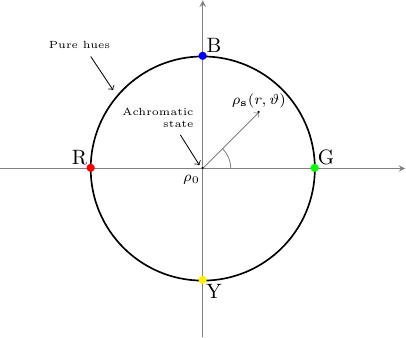}
	\caption{The Bloch disk. The colors chosen to visualize the intersections of the border with the axes will be motivated in section \ref{sec:applications}.}
	\label{fig:disk}
\end{figure}

It follows that the convex set of density matrices ${\cal S}(\R^2)$ of the rebit is parameterized by the points of $\mathcal D$, which, in quantum theories, is commonly referred to as the \textit{Bloch disk} and it is depicted in Figure \ref{fig:disk}. 
It can be seen that $\rho_0$ is parameterized by the center of the disk and $\rho(r,\vartheta)$ represents a generic density matrix expressed in polar coordinates, i.e.
\beq
\rho(r,\vartheta) = \frac{1}{2}\begin{pmatrix}
	1+r\cos \vartheta & r\sin \vartheta \\ r\sin \vartheta & 1-r\cos \vartheta
\end{pmatrix}, \quad r\in[0,1], \; \vartheta\in [0,2\pi).
\eeq

Since $d=\dim_{\mathbb R}(\h)=2$, all the orthogonal channels are exhausted by rotation matrices\footnote{As proven in \cite{Heinosaari:2011}, this holds only in dimension 2, so either for the rebit or the qubit.}.  Moreover, $N=2(2+1)/2=3$ and so the rotation matrix $R$ appearing in theorem \ref{th:roto} belongs to SO$(2)$ and labels orthogonal channels $\sigma_{\Omega_R}:{\cal S}(\R^2)\to {\cal S}(\R^2)$ such that
\beq\label{eq:fromRtoOmega}
\sigma_{\Omega_R}(\rho):=\Omega_R\, \rho \, {\Omega_R}^t,
\eeq 
with ${\Omega_R}^t=\Omega_R^{-1}$ and, thanks to eq. \eqref{eq:oro}, ${R}_{jk}={\rm Tr}(\sigma_j \, \Omega_R \, \sigma_k \, {\Omega_R}^t)/2$, for all $j,k=1,2$.

Finally, a generic channel $C$ induces an affine transformation of the Bloch vector $\textbf{v}^\rho\in \cal D$, so there exist $\textbf{w}\in \R^2$ and $A\in M(2,\R)$ such that 
\beq\label{eq:generic_chan}
C(\rho)=\half(I_2+(\textbf{w}+A\mathbf{v}^\rho)\cdot\vec{\sigma}),
\eeq 
or, if $C$ is unital,
\beq\label{eq:generic_unital_chan}
C(\rho)=\half(I_2+A\mathbf{v}^\rho\cdot\vec{\sigma}).
\eeq 

The information recalled above imply that we can classify the rebit channels by determining what are the possible matrices $A$ and vectors $\textbf{w}$ that may appear in eq. \eqref{eq:generic_chan} such that $C$ remains a channel. Of course, the classification would be greatly simplified if the matrix $A$ were diagonal or at least diagonalizable, however this may not be achievable. Instead, it is always possible to write a singular value decomposition of $A$:
\beq\label{eq:SVDA}
A=O_1\Sigma O^t_2,
\eeq 
where $O_1,O_2\in $ O$(2)$ and $\Sigma = {\rm diag}(\mu_1,\mu_2)$, where $\mu_1,\mu_2\in [0,+\infty)$ are the singular values of $A$.

Adapting the strategy used in \cite{Ruskai:2002} to the rebit case, we now show how it is possible to handle eq. \eqref{eq:SVDA} in order to understand how rebit channels act on the unit disk, up to a rotation, by replacing $A$ with a diagonal matrix $D$, originated from $\Sigma$, which is able to encompass also the general case in which the diagonal entries do not have the same sign.

We start by the basic remark that  $O_j\in$ O$(2)$ can be seen as the product of a rotation matrix $R_j\in$ SO$(2)$ with the matrix 
\beq 
\mathcal I=
\begin{cases} 
	{\rm diag}(1,1)\equiv I_2 & \text{ if } O_j\in \text{SO}(2)\\
	{\rm diag}(1,-1)\equiv I_2' & \text{ if } O_j\in \text{O}(2)\setminus \text{SO}(2)
\end{cases},
\eeq 
the second case representing a reflection around the horizontal axis. 

If both $O_1$ and $O_2$ belong to SO$(2)$ or to $\text{O}(2)\setminus \text{SO}(2)$ then we can rewrite  $O_j=R_j\cal I$ in eq. \eqref{eq:SVDA} where $\cal I$ is either the identity or the reflection, respectively. Taking into account that $\mathcal I^2=I_2$ in both cases and that $\Sigma$ and  $\mathcal I$ commute, we obtain
\beq 
A=R_1\mathcal I \Sigma \mathcal I R^t_2=R_1\Sigma R^t_2.
\eeq 
If, instead, $O_i\in$ SO$(2)$ and $O_j\in\text{O}(2)\setminus \text{SO}(2)$, with $i,j\in\{1,2\}$, $i\neq j$, then eq. \eqref{eq:SVDA} becomes
\beq 
A=R_1\Sigma' R^t_2, 
\eeq 
where $\Sigma':=I_2'\Sigma = \Sigma I_2'={\rm diag}(\mu_1,-\mu_2)$, which is \textit{no longer positive semi-definite}.

By introducing the decomposition of the identity $R_2R_2^t$ after $R_1$ in the two previous expressions of $A$, we get either $A=R_1 R_2^t R_2 \Sigma R_2^t$ or $A=R_1 R_2^t R_2 \Sigma' R_2^t$, but 
\beq R:=R_1 R_2^t \eeq 
belongs to SO$(2)$ and  
\beq S:= \begin{cases} R_2 \Sigma R_2^t \\
	\text{or} \\
	R_2 \Sigma' R_2^t
\end{cases} \eeq 
is such that $S^t=S$, i.e. it is a real symmetric matrix and, as such, diagonalizable. 

Notice that $S$ is not necessarily a positive semi-definite matrix since the conjugation of $\Sigma$ or $\Sigma'$ with the rotation matrix $R_2$ does not affect their eigenvalues and $\Sigma'$ is not positive-definite. Hence, if $\mathcal O\in$ O$(2)$ is the orthogonal matrix that diagonalizes $S$, then 
\beq\label{eq:Adecomprima}
A=R\mathcal O D\mathcal O^t,
\eeq 
where $D={\rm diag}(\lambda_1,\lambda_2)$, with  $\lambda_j\in \R$ and $|\lambda_j|=\mu_j$, $j=1,2$. Now we can repeat the same argument used above, i.e. we can write $\mathcal O=R_3 \mathcal I$, with $R_3\in$ SO$(2)$, so that eq. \eqref{eq:Adecomprima} becomes $A=RR_3 \mathcal I D \mathcal I R_3^t=RR_3 D R_3^t$. 

By defining $\mathcal R_1:=RR_3\in $ SO$(2)$ and $\mathcal R_2:=R_3^t\in $ SO$(2)$ we can finally write:
\beq\label{eq:Adecomp}
A=\mathcal R_1 D \mathcal R_2.
\eeq 
A natural question that arises now is that if the matrix factorization $A=\mathcal R_1 D \mathcal R_2$ implies an analogous channel factorization, the answer is positive thanks to the fact that $\mathcal R_1$ and $\mathcal R_2$ are special orthogonal matrices and we are working in dimension 2, so  they are automatically associated to orthogonal channels $\sigma_{\Omega_{\mathcal R_1}}$ and $\sigma_{\Omega_{\mathcal R_2}}$, which simply act on a Bloch vector via the matrix $\mathcal R_1$ and $\mathcal R_2$, respectively. So, the only degree of freedom that we have is in the definition of the channel corresponding to the matrix $D$. It is quite simple to recognize that the sequential transformation that must be applied to the Bloch vector $\textbf{v}^\rho$ in order to obtain formula \eqref{eq:generic_chan} with $A=\mathcal R_1 D \mathcal R_2$ is the following:
\beq \textbf{v}^\rho\stackrel{\sigma_{\Omega_{\mathcal R_2}}}{\mapsto}\mathcal R_2 \textbf{v}^\rho \stackrel{C_D}{\mapsto} \mathcal R_1^t \textbf{w} +D\mathcal R_2 \textbf{v}^\rho \stackrel{\sigma_{\Omega_{\mathcal R_1}}}{\mapsto} \mathcal R_1(\mathcal R_1^t \textbf{w} +D\mathcal R_2 \textbf{v}^\rho)=\textbf{w} +\mathcal R_1 D\mathcal R_2 \textbf{v}^\rho.\eeq 
The action of the second transformation on the density matrix $\rho$ is 
\beq\label{eq:CD}
C_{D}(\rho)=C_{D}\left(\half(I_2+\mathbf{v}^\rho\cdot\vec{\sigma})\right):=\begin{cases}
	\half\big(I_2+(\mathcal R_1^t\mathbf{w}+D\mathcal R_2\mathbf{v}^\rho)\cdot\vec{\sigma}\big) & \text{ if } C \text{ is not unital}\\
	\\
	\half(I_2+D\mathcal R_2\mathbf{v}^\rho\cdot\vec{\sigma})  & \text{ if } C \text{ is unital.}
\end{cases}
\eeq 
We can summarize what we have found in the following theorem.

\bt\label{th:decomp}
For every rebit channel $C$ there exist two rotation matrices $\mathcal R_1,\mathcal R_2\in $ \emph{SO}$(2)$ and a diagonal matrix $D={\rm diag}(\lambda_1,\lambda_2)$, with $\lambda_1,\lambda_2\in \R$, such that $C$ can be written as:
\beq\label{eq:decompC} 
C=  \sigma_{\Omega_{\mathcal R_1}} \circ C_{D} \circ \sigma_{\Omega_{\mathcal R_2}},
\eeq 
where $\sigma_{\Omega_{\mathcal R_j}}$ are the orthogonal channels associated to $\mathcal R_j$, $j=1,2$, and $C_{D}$ is defined as in eq. \eqref{eq:CD}.
\et 
As a consequence of this theorem, the rebit channels classification is equivalent to the determination of the constraints that must be satisfied by the map $C_D$ in order to be a channel. These constraints will clearly affect the shift vector $\textbf{w}$ and the entries of the diagonal matrix $D$.

The discussion of these constraints is long and quite technical and we postpone it to subsection \ref{subsec:analyticalconstraints}. Instead, in the next subsection we carry out the much simpler analysis of the possible geometric deformations of the unit disk $\cal D$, the state space of the rebit, induced by channels.

\subsection{Geometric deformation of the state space by rebit channels}

The action of the orthogonal channels corresponds simply to a rotation of the Bloch vector $\textbf{v}^\rho$ and the shift vector $\textbf{w}$. So, geometrically speaking, the only non-trivial part of the decomposition written in eq. \eqref{eq:decompC} is represented by action of $C_D$. 

To avoid an unnecessary complicated notation and keep the analysis as simple as possible, we will rewrite $\mathcal R_2 \textbf{v}^\rho$ and $\mathcal R_1^t \textbf{w}$ simply as $\textbf{v}^\rho=(v_1^\rho,v_2^\rho)$ and $\textbf{w}=(w_1,w_2)$, respectively. Using this simplified notation, we can write the modification of the vector $\mathbf{t}^\rho=\half(1,v_1^\rho,v_2^\rho)$ by $C_D$ as follows:
\beq\label{eq: action C_D}
\mathbf{t}^{C_D(\rho)}=\half\left(1,v_1^{C_D(\rho)},v_2^{C_D(\rho)}\right)=\half(1,w_1+\lambda_1v_1^\rho,w_2+\lambda_2v_2^\rho),\eeq
or
\beq\label{eq: action C_Dunital} \mathbf{t}^{C_D(\rho)}=\half\left(1,v_1^{C_D(\rho)},v_2^{C_D(\rho)}\right)=\half(1,\lambda_1v_1^\rho,\lambda_2v_2^\rho),
\eeq
in the case of unital channels. 

Let us analyze the latter case first: the action of $C_D$ on $\textbf{t}^\rho$ is represented in matrix form by
\beq \begin{pmatrix}
	1 & 0 & 0\\
	0 & \lambda_1 & 0\\
	0 & 0 & \lambda_2
\end{pmatrix}=\begin{pmatrix}
	1 & \mathbf{0}\\
	\mathbf{0} & D
\end{pmatrix},\eeq 
moreover, on one side we have that $\left(v_1^{C_D(\rho)}\right)^2+\left(v_2^{C_D(\rho)}\right)^2\le 1$ to assure the positivity of $C_D$, i.e. that $\textbf{v}^{C_D(\rho)}$ is still a state, and, on the other side, from \eqref{eq: action C_Dunital} we have that $v_j={v_j^{C_D(\rho)}}/{\lambda_j}$, $j=1,2$, so the constraint  $({v}^\rho_1)^2+({v}^\rho_2)^2\leq1$ is translated into the inequality  
\beq\label{eq: ellipse}\left(\frac{{v}^{C_D(\rho)}_1}{\lambda_1}\right)^2+\left(\frac{{v}^{C_D(\rho)}_2}{\lambda_2}\right)^2\leq1.\eeq
This means that $C_D$ maps the set of rebit states $\cal D$ in the two dimensional closed ellipsoid embedded in $\cal D$ represented by equation \eqref{eq: ellipse}. On its border we find the images by $C_D$ of the pure states of the rebit, parametrized by the elements of the unit circle. Since $\lambda_1$ and $\lambda_2$ are the semi-axes of this ellipsoid, they must verify 
\beq |\lambda_j|\le 1, \; j=1,2.\eeq  
The centre of the Bloch disk remains fixed by the action of $C_D$ only if $C$ is unital. In the more general case of a non-unital channel $C$ we have, by eq. \eqref{eq: action C_D}, that the action of $C_D$ on $\textbf{t}^\rho$ can be represented via the affine matrix 
\beq\label{eq: Diagonal A rebit}
\mathbb A =\begin{pmatrix}
	1 & 0 & 0\\
	w_1 & \lambda_1 & 0\\
	w_2 & 0 & \lambda_2
\end{pmatrix}=\begin{pmatrix}
	1 & \mathbf{0}\\
	\mathbf{w} & D
\end{pmatrix}\eeq 
and the image of the Bloch disk under the action of $C_D$ is an ellipsoid with origin of the semi-axes given by $(w_1,w_2)$ and with analytic expression given by:
\beq\label{eq: translated ellipse}\left(\frac{{v}^{C_D(\rho)}_1-w_1}{\lambda_1}\right)^2+\left(\frac{{v}^{C_D(\rho)}_2-w_2}{\lambda_2}\right)^2\leq1,
\eeq 
Again, to guarantee the positivity of $C_D$, i.e. that the constraints defining states are still valid, we must have
\beq |w_j|\le 1, \; j=1,2. \eeq
$\lambda_1$ and $\lambda_2$ will be called \textit{scale coefficients}, while we will refer to $w_1$ and $w_2$ as \textit{shift coefficients}. 

\subsection{Analytical constraints to be satisfied by $C_D$ to be a channel}\label{subsec:analyticalconstraints}

In order to be a channel, $C_D$ must be completely positive and trace-preserving. This last property is easily seen to be satisfied, in fact the first row of the matrix appearing in formula \eqref{eq: Diagonal A rebit} is $(1,0,0)$, hence it leaves the first component of the vector $\textbf{t}^\rho$ unaltered. In section \ref{sec:mathpreliminars} we have seen that this implies that $C_D$ is trace-preserving.  So, only complete positivity must be examined.

In \cite{Ruskai:2002}, necessary and sufficient conditions for the complete positivity in the qubit case have been established by using Choi's theorem. For that, it is essential to rewrite the four matrices $E_{jk}$, $j,k=1,2$, of the canonical basis of $M(2,\C)$ as a linear combination of the identity plus the \textit{three} Pauli matrices, including the one with complex entries.  However, in the rebit case we have at disposal only the identity plus the \textit{two} real Pauli matrices and so this technique cannot be pursued any longer. Hence, other strategies must be considered.

In what follows we will show that the $\chi$-matrix representation recalled in section \ref{sec:mathpreliminars} can be used instead of the Choi theorem to establish necessary and sufficient conditions for the complete positivity of the map $C_D$.

In order to build the $\chi$-matrix associated to $C_D$ it is convenient to fix the orthonormal Bloch basis of $\h(2,\R)$, i.e. $(\sigma_0, \sigma_1, \sigma_2)/\sqrt{2}$. Thanks to eqs. \eqref{eq: coefficients chi} and \eqref{eq: Diagonal A rebit} we get
\begin{equation}
	\begin{split}
		\chi_{r,s}&=\frac{1}{4}\sum_{k=0}^2{\rm Tr}[\sigma_s\sigma_k\sigma_rC_D(\sigma_k)]\\
		&=\frac{1}{4}\Big[{\rm Tr}\big(\sigma_s \sigma_r(I_2+w_1\sigma_1+w_2\sigma_2)\big)+\lambda_1 {\rm Tr}\big(\sigma_s \sigma_1\sigma_r\sigma_1\big)+\lambda_2{\rm Tr}\big(\sigma_s \sigma_2\sigma_r\sigma_2\big)\Big].
	\end{split}
\end{equation}
The straightforward computation of the matrix elements gives
\beq\label{eq: chi C}
\chi_{C_D}(\lambda_1,\lambda_2,w_1,w_2)=\half\pmat{1+\lambda_1+\lambda_2 & w_1 & w_2\\w_1 & 1+\lambda_1-\lambda_2 &0\\ w_2 & 0 & 1-\lambda_1+\lambda_2}.
\eeq
As recalled in section \ref{sec:mathpreliminars}, the complete positivity of $C_D$ is equivalent to the positive semi-definiteness of $\chi_{C_D}$. Being a real symmetric matrix, its eigenvalues are all real and so $\chi_{C_D}$ is positive semi-definite if and only if all its eigenvalues are non-negative. We will conduct the analysis of positive semi-definiteness following an increasing order of complexity. 

\subsubsection{The unital case ($w_1=w_2=0$)}
Unital rebit channels leave the centre of the Bloch disk invariant, so they have a null shift vector $\mathbf{w}=(w_1,w_2)=\mathbf{0}$.

\begin{figure}[!ht]
	\centering
	\includegraphics[width=2.5in]{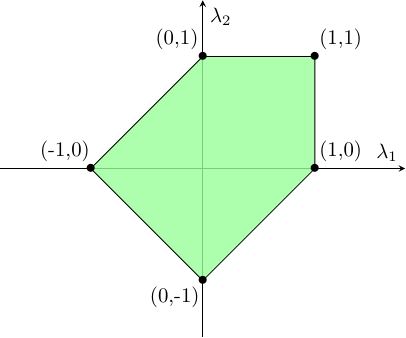}
	\caption{Admissibility region $\cal P$ for the parameters $\lambda_1$ and $\lambda_2$ to guarantee the complete positiveness of the map $C_D$ in the unital case.}
	\label{fig:rebit complete positivity region}
\end{figure}

This means that the $\chi-$matrix in this case reduces to the following diagonal form
$$\chi_{C_D}(\lambda_1,\lambda_2)=\half\pmat{1+\lambda_1+\lambda_2 & 0 & 0\\ 0 & 1+\lambda_1-\lambda_2 & 0\\ 0 & 0 & 1-\lambda_1+\lambda_2}.$$
Its eigenvalues are therefore its three diagonal elements, i.e.
\beq\label{eq: q0 q1 q2}
q_0:=\half(1+\lambda_1+\lambda_2), \quad
q_1:=\half(1+\lambda_1-\lambda_2), \quad
q_2:=\half(1-\lambda_1+\lambda_2)
\eeq
and so, $C_D$ is completely positive if and only if 
\beq\label{eq: cases lambda}
\begin{cases}
	\lambda_2\ge-1\pm\lambda_1\\
	\lambda_2\le1+\lambda_1\\
\end{cases}.
\eeq
Since $(\lambda_1,\lambda_2)\in [-1,1]\times [-1,1]\subset \R^2$, the three inequalities appearing in the system \eqref{eq: cases lambda} define the admissibility region $\cal P$ for the parameters, which lies inside the pentagon of Figure \ref{fig:rebit complete positivity region}. 

\subsubsection{The non-unital cases with either $w_1=0$ or $w_2=0$}
We start by considering $\mathbf{w}=(0,w_2)$, $|w_2|\le 1$, $w_2\neq 0$. This means that the shift occurs only in the direction of the $\lambda_2-$axis. In this case the $\chi$-matrix is 
$$\chi_{C_D}(\lambda_1,\lambda_2,w_2)=\half\pmat{1+\lambda_1 +\lambda_2 &0 &  w_2 \\
	0 & 1+\lambda_1-\lambda_2 & 0\\
	w_2 & 0 & 1-\lambda_1+\lambda_2}.$$
First of all we notice that the condition $q_j\ge0$ for $j=0,1,2$ is still a necessary condition for positive semi-definiteness of the matrix and hence for complete positivity of $C_D$. This can be seen by selecting the vectors $e_j$ of the canonical basis of $\R^3$ and performing the scalar product $e_j^t \chi_{C_D}(\lambda_1,\lambda_2,w_2) e_j$, which reproduces exactly the values $q_j$, $j=1,2,3$.  

In order to give also sufficient conditions we compute the eigenvalues of this matrix. The characteristic polynomial of $\chi_{C_D}$ is:
\beq
\footnotesize{P(x)=\left[\half(1+\lambda_1-\lambda_2)-x\right]\left[x-\half\left(1+\lambda_2+\sqrt{\lambda_1^2+w_2^2}\right)\right]\left[x-\half\left(1+\lambda_2-\sqrt{\lambda_1^2+w_2^2}\right)\right],}
\eeq
and therefore the three eigenvalues are 
$$
\begin{cases}
	\mu_0:=\half(1+\lambda_1-\lambda_2)\\
	\mu_{\pm}:=\half\left(1+\lambda_2\pm\sqrt{\lambda_1^2+w_2^2}\right).
\end{cases}
$$
The condition $\mu_0\ge0$ is equivalent to the necessary condition $q_1\ge0$ and the request that $\mu_+\ge0$ gives $\lambda_2\geq-\sqrt{\lambda_1^2+w_2^2}-1$, which is always true, since it is necessary that $|\lambda_j|\le1$ and $|w_j|\le1$ for all $j=1,2$ in order to guarantee positivity.

Finally, $\mu_{-}\geq0$ gives $1+\lambda_2\geq\sqrt{\lambda_1^2+w_2^2}$, which, squaring both sides gives an upper bound on $w_2$, i.e. on the possible vertical shifts of the ellipsoid, precisely:
\beq\label{eq: complete positivity 1}
w_2^2\leq(\lambda_2+1)^2-\lambda_1^2.
\eeq

The case in which $\mathbf{w}=(w_1,0)$, i.e. when the shift occurs only in the $\lambda_1-$axis, is analogous to the case described above. Proceeding as before, we obtain the following upper bound on the possible horizontal shifts of the ellipsoid:
\beq\label{eq: complete positivity 2}
w_1^2\leq(\lambda_1+1)^2-\lambda_2^2.
\eeq

\subsubsection{The non-unital case with a generic $\mathbf{w}=(w_1,w_2)$}
In the general case, even if it is possible to compute the eigenvalues of $\chi_{C_D}$, their explicit expression is so complicated that their non-negativity cannot be studied analytically. We can avoid this problem by studying the characteristic polynomial of the expression of $\chi_{C_D}$ in eq. \eqref{eq: chi C}, i.e.
\beq\label{eq: charpoly}
P(x)=-x^3+\frac{a}{2}x^2-\frac{b}{4}x+\det{(\chi_{C_D})},
\eeq
with $a=3+\lambda_1+\lambda_2$, $b=3 - (w_1^2 + w_2^2) + 2 (\lambda_1 + \lambda_2) - (\lambda_1+\lambda_2)^2$ and 
\beq
\det(\chi_{C_D})=\frac{1}{8}\Big[(1-\lambda_1+\lambda_2)\left[(1+\lambda_1+\lambda_2)(1+\lambda_1-\lambda_2)-w_1^2\right]-w_2^2(1+\lambda_1-\lambda_2)\Big].
\eeq
By Descartes' rule of signs, all the three roots of $P$ are positive if and only if the coefficients of the powers of $x$ present three variations, i.e. their signs change from one to the next. This happens if and only if $a,b$ and $\det{(\chi_{C_D})}$ are all $\ge0$. 

Since the condition $|\lambda_j|\le1$ must hold to guarantee positivity, $a\geq 0$ always holds. The request $b\ge0$ gives an upper bound on the norm of $\mathbf{w}$, that is 
\beq\label{eq: b>=0}
\|\mathbf{w}\|^2=w_1^2+w_2^2\le3+2(\lambda_1+\lambda_2)-(\lambda_1+\lambda_2)^2.
\eeq
Finally, $\det{(\chi_{C_D})}\ge0$ holds if and only if 
\begin{equation}\label{eq: det. condition}
	w_1^2(1-\lambda_1+\lambda_2)+w_2^2(1+\lambda_1-\lambda_2)\leq(1+\lambda_1+\lambda_2)(1+\lambda_1-\lambda_2)(1-\lambda_1+\lambda_2)=8\, q_0q_1q_2,
\end{equation}
having used definition \eqref{eq: q0 q1 q2} in the last equality.
The condition $\det(\chi_{C_D})\ge0$ always implies $b\ge0$. Indeed, the region described by \eqref{eq: b>=0} is a disk w.r.t. the two variables $(w_1,w_2)$, moreover
\beq 
\max_{|\lambda_{j}|\le1, \, j=1,2}\{3+2(\lambda_1+\lambda_2)-(\lambda_1+\lambda_2)^2\}=4,
\eeq 
so the ray of this disk is always $\le 2$. On the other hand, when $1\pm\lambda_1+\lambda_2>0$ and $1+\lambda_1-\lambda_2>0$, the region described by \eqref{eq: det. condition} w.r.t $(w_1,w_2)$ is an ellipsoid that can be explicitly written as
\beq\label{eq: det. condition 5}
\frac{w_1^2}{\dfrac{1}{4}(1+\lambda_1+\lambda_2)(1+\lambda_1-\lambda_2)}+\frac{w_2^2}{\dfrac{1}{4}(1+\lambda_1+\lambda_2)(1-\lambda_1+\lambda_2)}\le4.
\eeq
Therefore, since 
\beq 
\max_{|\lambda_{j}|\le1, \, j=1,2}\{(1+\lambda_1+\lambda_2)(1+\lambda_1-\lambda_2)\}=\max_{|\lambda_{j}|\le1, \, j=1,2}\{(1+\lambda_1+\lambda_2)(1-\lambda_1+\lambda_2)\}=4,
\eeq  
it follows that both denominators of the inequality \eqref{eq: det. condition 5} are bounded by $1$ when $|\lambda_{j}|\le1$, $j=1,2$. This means that this ellipsoid described by the condition \eqref{eq: det. condition 5} is always contained into the disk described by eq. \eqref{eq: b>=0}. Moreover, inequality \eqref{eq: det. condition 5} gives a nice geometric characterization of condition \eqref{eq: det. condition}, saying that $C_D$ is completely positive if and only if its shift vector $\mathbf{w}$ belongs to the ellipsoid described by \eqref{eq: det. condition 5}.

Notice that when $(1-\lambda_1+\lambda_2)=0$ (or $(1+\lambda_1-\lambda_2)=0$, or $(1+\lambda_1+\lambda_2)=0$, respectively), inequality \eqref{eq: det. condition} forces $w_2=0$ (or $w_1=0$, or $w_1=w_2=0$, respectively) and hence the ellipsoid described by inequality \eqref{eq: det. condition} degenerates to a segment or to the origin $(w_1,w_2)=\mathbf{0}$ which are always contained into the disk described by eq. \eqref{eq: b>=0}.

In conclusion, $\det{(\chi_{C_D})}\ge0$ implies $b\ge0$ in every case, and hence the inequality \eqref{eq: det. condition} is necessary and sufficient to guarantee that the eigenvalues of $\chi_{C_D}$ are $\ge 0$. 

\subsection{The final classification theorem of rebit channels}

We summarize the analysis conducted above in the following theorem.

\bt\label{th:rebitclassification}
Every rebit channel $C$ can be decomposed as $C=  \sigma_{\Omega_{\mathcal R_1}} \circ C_{D} \circ \sigma_{\Omega_{\mathcal R_2}}$, where $\sigma_{\Omega_{\mathcal R_j}}$
are orthogonal channels associated to the rotation matrices $\mathcal R_j\in $ \emph{SO}$(2)$, $j=1,2$,  and $C_{D}$ can be represented in the Bloch basis $(\sigma_0,\sigma_1,\sigma_2)$ by the affine matrix
$$
\mathbb A =\begin{pmatrix}
	1 & 0 & 0\\
	w_1 & \lambda_1 & 0\\
	w_2 & 0 & \lambda_2
\end{pmatrix}=\begin{pmatrix}
	1 & \mathbf{0}\\
	\mathbf{w} & D
\end{pmatrix},
$$
with the coefficients $\lambda_j,w_j$, $j=1,2$, satisfying the following conditions:
\beq
\begin{cases}
	q_1=1+\lambda_1+\lambda_2 \ge 0\\
	q_2=1+\lambda_1-\lambda_2 \ge 0\\
	q_3=1-\lambda_1+\lambda_2 \ge 0\\
	\frac{w_1^2}{(1+\lambda_1+\lambda_2)(1+\lambda_1-\lambda_2)}+\frac{w_2^2}{(1+\lambda_1+\lambda_2)(1-\lambda_1+\lambda_2)}\le1
\end{cases}.
\eeq
\et
The condition $\lambda_1=\lambda_2=1$ describes the identity channel, in this case ${\bf w}$ must be null to preserve the Bloch disk.

Geometrically, because of the presence of $C_D$, the action of $C$ on the Bloch disk representing the state space of the rebit is to shrink it to an ellipsoidal region contained in the unit disk in $\R^2$, which can also degenerate into a line segment or in a point. Either the semi-axes of the ellipsoidal region or the line segment can be tilted with respect to the coordinate axes due to the presence of the orthogonal channels $\sigma_{\Omega_{\mathcal R_j}}$, $j=1,2$.


\section{Noticeable rebit channels}\label{sec:noticeablechannels}

In the previous section we have seen that the really interesting information about the deformation of the Bloch disk by a unital rebit channel is contained in way the scaling parameters $\lambda_1,\lambda_2$ vary inside the admissible region $\cal P$ depicted in Figure \ref{fig:rebit complete positivity region} and described by inequalities \eqref{eq: cases lambda}. In fact, the presence of the orthogonal channels appearing in theorem \ref{th:rebitclassification} simply amounts to a rotation of the Bloch vectors. So, in what follows, we will just concentrate on the diagonal unital channel $C_D$, recalling that its action on a general state $\rho=\half(\sigma_0+r\cos(\vartheta)\sigma_1+r\sin(\vartheta)\sigma_2)$ is the following:
\beq\label{eq:actunit}
\pmat{1 & 0 & 0\\ 0 & \lambda_1 & 0\\0 & 0 & \lambda_2}\half\pmat{1\\ r\cos(\vartheta)\\ r\sin(\vartheta)}=\half\pmat{1\\ \lambda_1 r\cos(\vartheta)\\ \lambda_2 r\sin(\vartheta)}.
\eeq 
Let us start by the characterisation of the so-called \textit{Kraus rank} of $C_D$, i.e. the number of non-zero eigenvalues of the associated $\chi-$matrix:
\begin{itemize}
	\item the vertices of $\cal P$ given by  $(\lambda_1,\lambda_2)=(-1,0)$ and  $(\lambda_1,\lambda_2)=(0,-1)$ correspond to \textit{rank-1} channels because $q_1=q_2=0,q_3=2$ and $q_1=q_3=0,q_2=2$, respectively;
	\item the points belonging to the three diagonal edges of $\cal P$ defined by the conditions $\lambda_2=\pm\lambda_1-1$ and $\lambda_2=\lambda_1+1$ correspond to \textit{rank-2} channels. This is straightforward to see, since a single vanishing eigenvalue $q_j$ in eq. \eqref{eq: q0 q1 q2} defines a diagonal edge, and so the intersection of two diagonal edges correspond to two vanishing eigenvalues;
	\item all the other points of $\cal P$ correspond to \textit{full-rank} (i.e. rank-3) channels.
\end{itemize}
Now we pass to the analysis of some noticeable rebit channel.
\begin{enumerate}
	\item The \textit{identity channel}, which correspond to $(\lambda_1,\lambda_2)=(1,1)$, leaves the Bloch disk invariant. This is the only channel with this behaviour. In fact, apart from the case just examined, the length of the disk axes could be preserved only by the couples  $(-1,-1), (1,-1), (-1,1)$ which, however, lie outside the admissibility region $\cal P$.
	\item The \textit{phase flip channels}: by substituting $(\lambda_1,\lambda_2)=(1-p,1)$, with $0<p<1$, in eq. \eqref{eq:actunit}, the corresponding channels transform the Bloch disk into an ellipsoid that intersects it at the poles, i.e. the yellow and blue points of Figure \ref{fig:disk}, which remain fixed, while the disk is shrunk in the horizontal direction. The case $p=1$, i.e. $(\lambda_1,\lambda_2)=(0,1)$, corresponds to a rank-2 channel which reduces the Bloch disk to the vertical axis.
	
	The case $(\lambda_1,\lambda_2)=(1,1-p)$ is exactly analogous, with the role of the two axes swapped. When $p=1$ the Bloch disk degenerates into the horizontal axis. Figure \ref{fig: ex. rebit} illustrates all these cases.
	\begin{figure}[!ht]
		\centering
		\begin{minipage}[c]{\textwidth}
			\centering 
			\subfigure[]
			{\includegraphics[scale=0.45]{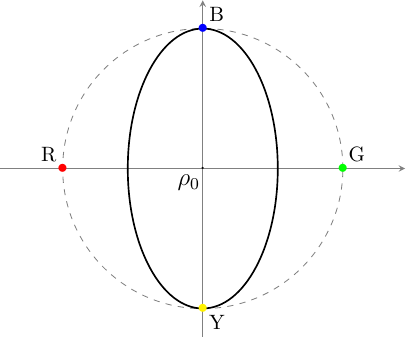}}
			\hspace{2mm}
			\subfigure[]
			{\includegraphics[scale=0.45]{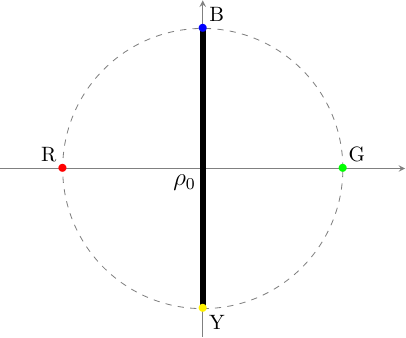}}
			\hspace{2mm}
			\subfigure[]
			{\includegraphics[scale=0.45]{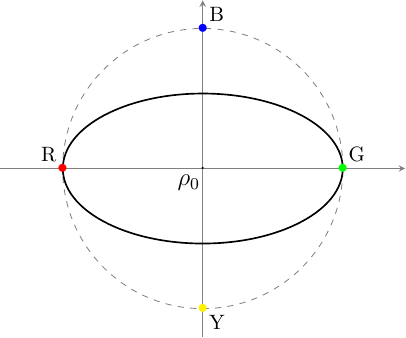}}
			\hspace{2mm}
			\subfigure[]
			{\includegraphics[scale=0.45]{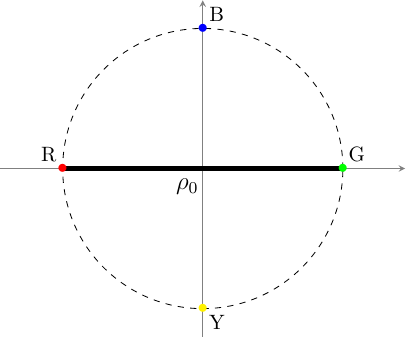}}
			\caption{$(\lambda_1,\lambda_2)$ for phase flip channels. (a): $(1-p,1)$, (b): $(0,1)$, (c): $(1,1-p)$, (d): $(1,0)$, $p\in (0,1)$.}
			\label{fig: ex. rebit}
		\end{minipage}
	\end{figure}

	\item \textit{Depolarizing channels} are channel such that $|\lambda_1|=|\lambda_2|$, i.e. corresponding to points lying on the two bisectors of $\cal P$ in Figure \ref{fig:rebit complete positivity region}. These channels simply shrink the Bloch disk to another disk with radius $r<1$. If one of the $\lambda_j$'s is negative, the channel reflects also the disk with respect to one of the two (or both) axes. When $\lambda_1=\lambda_2=0$, the entire disk degenerates to the single central point which represents the maximally mixed state $\rho_0$. This is the so-called \textit{completely depolarizing channel}.
	\item \textit{Linear channels} are defined by  the condition that either $\lambda_1$ or $\lambda_2$ or both are equal to $0$, i.e. they correspond to points lying on the horizontal or vertical axis in Figure \ref{fig:rebit complete positivity region}. When $(\lambda_1,\lambda_2)=(q,0)$ (or $(\lambda_1,\lambda_2)=(0,q)$, respectively), with $-1<q<1$, the entire Bloch disk is sent to an horizontal (or vertical, respectively) segment of length $2|q|$. When $q<0$ also a reflection is applied. When $q=0$ we recover the completely depolarizing channel.
	
	When $q=-1$ we have the two rank-1 channels corresponding to the vertices $(0,-1)$ and $(-1,0)$, which are the intersection of two of the diagonal edges. These channels first shrink the Bloch disk to a segment as in case (ii), but in addition they reflect it with respect to the horizontal or vertical axis, respectively.
	\item All the other channels correspond to points lying either on the ‘diagonal' edges or to the interior of Figure \ref{fig:rebit complete positivity region}. In the first case we have rank-2 channels, in the second one full-rank channels. All the corresponding channels transform the Bloch disk to a generic ellipsoid. 
\end{enumerate}

\begin{figure}[ht!]
	\centering
	\includegraphics[width=4in]{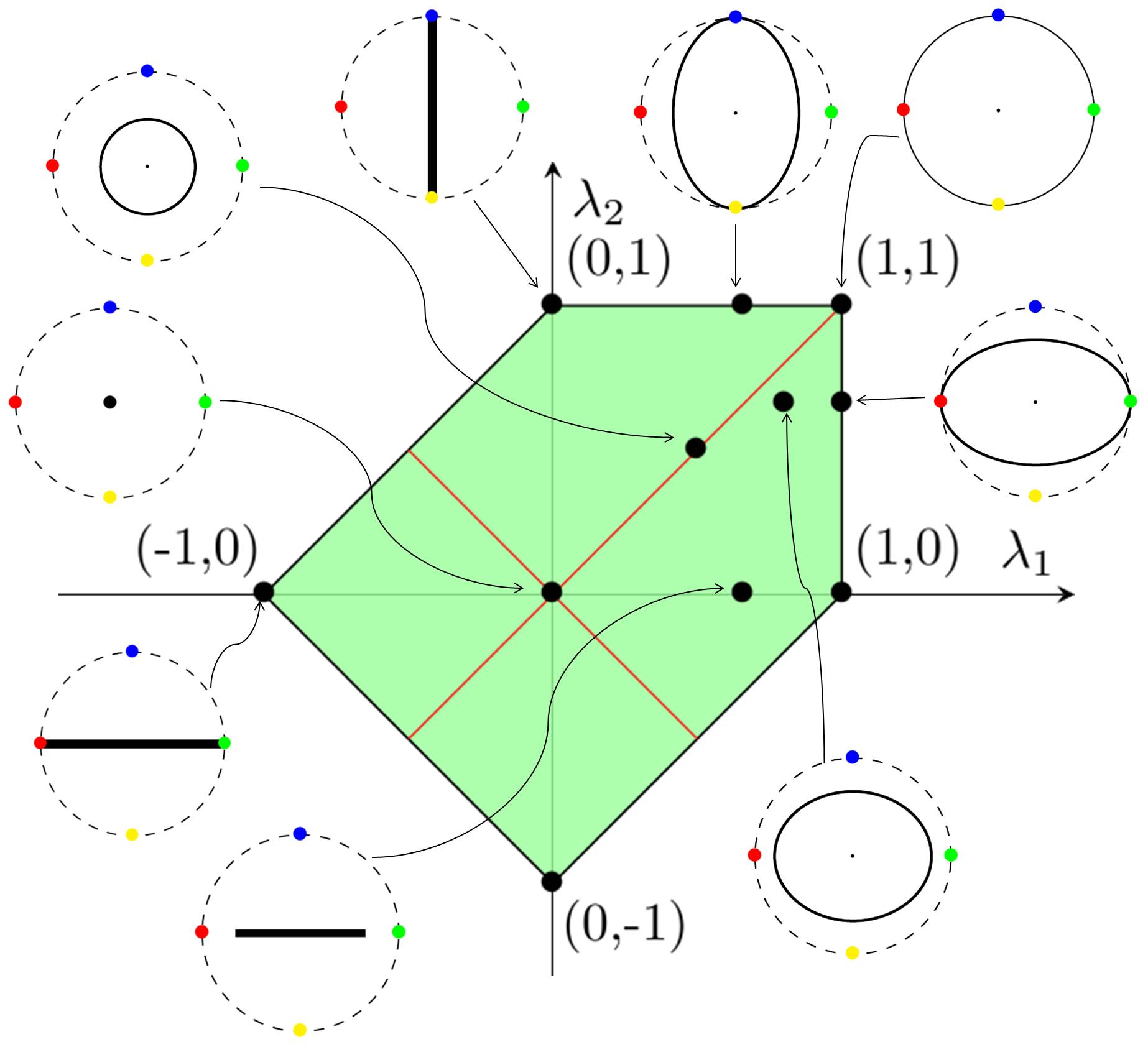}
	\caption{A visual representation of the cases described in the previous list. Starting from the identity and proceeding in a clockwise ordering we have: a phase flip of the type $(\lambda_1,\lambda_2)=(1,1-p)$, a general case, a linear channel, a rank-1 degenerate channel, the completely depolarizing channel, a depolarizing channel, a degenerate phase flip and finally a phase flip of the type $(\lambda_1,\lambda_2)=(1-p,1)$.}\label{fig:recap_unital_channels}
\end{figure}

Figure \ref{fig:recap_unital_channels} gives a compact illustration of the channels discussed in the previous list depending on the position of the couple of scaling parameters $(\lambda_1,\lambda_2)\in \cal P$. Due to Theorem \ref{th:decomp}, the previously discussed transformations of the Bloch disk must be considered up to rotations. 

\section{Future applications of the classification theorem}\label{sec:applications}

Rebit properties cannot be deduced from those of a qubit simply by replacing the complex field $\C$ with the real field $\R$. For instance, Wootters has proven that, while local measurements are enough to determine the state of an open qubit system, this is not necessarily true for a rebit, see  \cite{Wootters:1990}. This is just one of the peculiar features of rebits that make their analysis worthwhile. In this section we would like to briefly discuss two open problems that may benefit from the rebit channel classification theorem. 

The first is related to rebit dynamics: for a closed (complex) quantum system with Hamiltonian $H$, the unitary evolution of density matrices is governed by the Schrödinger-Liouville equation $\dot \rho=-i[H,\rho]$. Its non-unitary counterpart for open quantum systems in the Markovian approximation is the Lindblad master equation, which still involves the imaginary unit, see e.g. \cite{Auletta:09}. This raises the natural question about how the dynamics of real quantum systems can be described. At this moment, the question remains a much debated and interesting open problem, see e.g. \cite{McKague:2009,Aleksandrova:13,Moretti:17,Renou:2021,Fuchs:2022}. If one wished to analyze the dynamical change of rebit states in terms of channels, it seems reasonable to assume that having available the complete classification of these maps would be of great benefit. 

The second application refers to a recent model of human color perception formulated in the framework of quantum information, see \cite{BerthierProvenzi:2022PRS,Berthier:22SIAM} for the two more recent and complete publications about this topic. In such a model, human observers  perform perceptual measurements on visual scenes prepared in so-called \textit{chromatic states}, which are nothing but states of a particular rebit system called Hering's rebit \cite{Berthier:2020}. This name honors the famous Hering's theory in which colors are described by means of two degrees of  incompatible chromatic sensations: red-green and yellow-blue. In the Bloch disk, the opponent colors are antipodal, which explains their placement in the depiction of the Bloch disk in Figure \ref{fig:disk}.

The crucial fact is that a perceptual measurement is modeled via an effect associated to a human observer. The measurement modifies the initial chromatic state via a Lüders operation, giving rise to a post-measurement generalized state. It is well-known that color perception abilities vary from person to person and that some individuals may also be affected by different degrees and types of color perception anomalies. For this reason, an important open problem in colorimetry is the determination of the so-called \textit{individual chromatic space} $\cal I$, i.e. the space of chromatic sensations that can actually be perceivable by a given observer. In the quantum model, $\cal I$ is a convex subset of the Bloch disk. In a future paper, we will explore the possibility of identifying the individual color space with the image of the Bloch disk via a rebit channel uniquely associated to a human observer. For this purpose, the classification theorem proven in this paper will turn out to be of fundamental importance.

\section{Conclusions and perspectives}\label{sec:conclusions}

We have detailed the complete classification of rebit channels and underlined that the strategy used in the 2002 paper  \cite{Ruskai:2002} for the classification of qubit channels, founded on Choi's theorem, must be replaced by an alternative procedure based on the $\chi$-matrix representation. 

The classification theorem has not only a theoretical significance, but it may help understanding the dynamical evolution of real quantum systems and lead to important applications in a recent quantum information-based color perception model which will be explored in future contributions.

\bibliographystyle{plain} 
\bibliography{bibliography}

\end{document}